\documentclass[conference,letterpaper]{IEEEtran}
\IEEEoverridecommandlockouts
\usepackage{amsmath,amssymb,amsfonts}
\usepackage{algorithmic}
\usepackage{graphicx}
\usepackage{textcomp}
\usepackage{xcolor}
\usepackage[style=ieee, maxnames=6, minnames=1, doi=false, isbn=false, url=false]{biblatex}
\usepackage{threeparttable}
\usepackage{subcaption}
\usepackage{multirow, multicol}
\usepackage{threeparttable}
\usepackage{booktabs}
\def\BibTeX{{\rm B\kern-.05em{\sc i\kern-.025em b}\kern-.08em
    T\kern-.1667em\lower.7ex\hbox{E}\kern-.125emX}}
\begin{document}

\title{VFocus: Better Verilog Generation from Large Language Model via Focused Reasoning\\
}

\author{
\IEEEauthorblockN{Zhuorui Zhao\textsuperscript{1}, Bing Li\textsuperscript{2}, Grace Li Zhang\textsuperscript{3}, Ulf Schlichtmann\textsuperscript{1}}
\IEEEauthorblockA{\textsuperscript{1}\textit{Chair of Electronic Design Automation, Technical University of Munich (TUM)} \\
\textsuperscript{2}\textit{Digital Integrated Systems Group, University of Siegen} \\
\textsuperscript{3}\textit{Hardware for Artificial Intelligence Group, Technical University of Darmstadt} \\
Email: $\{$zhuorui.zhao, ulf.schlichtmann$\}$@tum.de, bing.li@uni-siegen.de, grace.zhang@tu-darmstadt.de}
}

\maketitle

\begin{abstract} 
Large Language Models (LLMs) have shown impressive potential in generating Verilog codes, but ensuring functional correctness remains a challenge. Existing approaches often rely on self-consistency or simulation feedback to select the best candidate, but they miss opportunities to focus LLM reasoning on the most informative parts of the design. We propose VFocus, a three-stage framework that enhances Verilog generation by sharpening the focus of LLM reasoning onto critical decision points in the code generation process. In the \textbf{pre-ranking stage}, VFocus generates multiple code candidates through LLM prompting, retries for syntactically valid outputs, and introduces a \textit{Density-guided Filtering} to retain candidates that fall within the “reasoning sweet spot” for functional correctness. In the \textbf{ranking stage}, we simulate each code candidate using an automatically generated testbench and apply self-consistency-based clustering to identify the most consistent outputs. Finally, in the \textbf{post-ranking refinement stage}, VFocus performs inconsistency mining on top-ranked candidates and invokes reasoning-augmented LLM prompts for candidate refinement. Experiments on the VerilogEval-Human benchmark show that VFocus significantly improves the pass@1 correctness across multiple reasoning LLMs, demonstrating its effectiveness in enhancing Verilog generation for complex hardware design tasks.
\end{abstract}


\section{Introduction}
\label{sec:Introduction}

Hardware design is becoming increasingly important in today’s computing landscape, driven by the demand for customized accelerators, domain-specific architectures, and efficient SoC solutions. As hardware complexity grows, automating the generation of hardware description languages (HDLs) such as Verilog has emerged as a promising way to reduce development costs and accelerate innovation.

Large Language Models (LLMs) have shown remarkable performance in generating software code and are now being explored for hardware design tasks. While LLMs have demonstrated encouraging results in HDL generation from natural language specifications, their accuracy still lags behind software code generation, mainly due to the limited availability of high-quality HDL datasets. 

Current approaches to improve HDL generation fall into two categories: training-based and training-free methods.
Training-based approaches attempt to fine-tune general-purpose code LLMs on Verilog-specific datasets~\cite{pei_betterv_2024}\cite{lu_rtllm_2024}\cite{zhao_codev_2024}. However, due to data scarcity and domain complexity, these fine-tuning attempts gain less improvement compared to model scaling and training paradigm shift in general-purpose LLMs~\cite{pan_survey_2025}.
Alternatively, training-free methods aim to enhance code generation at inference time through workflows or agentic pipelines. The workflows involves prompt engineering that breaks down tasks into predefined substeps~\cite{sun_paradigm-based_2025}. The agentic pipelines often leverage testbenches for feedback~\cite{zhao_mage_2024}\cite{ho_verilogcoder_2025}. These techniques typically require human-written testbenches or depend on human-in-the-loop debugging, which limits scalability. However, while automatic testbench generation by LLMs is feasible, it tends to be less reliable than direct code generation, making testbench-code co-generation challenging~\cite{qiu_autobench_2024}.
Another training-free method, VRank~\cite{zhao_vrank_2025} leverages self-consistency to select high-quality Verilog code. Instead of trusting the correctness of generated testbenches, VRank uses them only as a proxy to measure consistency across multiple generated code samples. By identifying the candidate that aligns best with the majority outcome of testbench simulations, VRank significantly improves functional correctness without requiring human feedback.



In this work, we present VFocus, a reasoning-enhanced framework that is training-free, does not depend on reliable test benches, and addresses problems in previous work, such as~\cite{zhao_vrank_2025} by leveraging the reasoning ability of LLM. VFocus performs both pre-ranking sampling enhancement and post-ranking refinement of generated Verilog codes. Through inconsistency mining and reasoning-guided improvement, our method not only selects better codes but also improves their correctness. 

The contributions of our paper are summarized as follows:

\begin{itemize}
    \item We propose VFocus, a three-stage Verilog code generation framework that enhances reasoning LLMs by sharpening their focus on critical decision points during code generation and refinement.
    \item In the pre-ranking stage, we introduce a novel Density-guided Filtering method and validity check of code samples to retain candidates that fall within the optimal complexity range (“reasoning sweet spot”) for functional correctness, thus improving sample quality before selection.
    \item In the post-ranking refinement stage, we perform inconsistency mining to identify behavioral uncertainty in top-ranked candidates and use reasoning-augmented prompts to refine codes automatically.
   \item Experimental results show that VFocus achieves a significant improvement in pass@1 rates on reasoning LLMs, which highlights the potential of our approach to advance automated Verilog code generation via scenario reasoning and sample selection.
\end{itemize}

The rest of the paper is organized as follows. Section \ref{sec:BackgroundandMotivation} presents an overview of existing work on Verilog generation and the related methods for language model reasoning. Section \ref{sec:Proposed Framework} explains our VFocus in detail. The experimental setup is explained in Section \ref{sec:ExperimentalSetup}. Section \ref{sec:ExperimentalResults} provides the research question and experimental results. Section \ref{sec:conclusion} concludes our work.

\section{Background and Motivation}
\label{sec:BackgroundandMotivation}

\subsection{training-free LLM frameworks for Verilog Generation}
\begin{figure}
    \centering
    \includegraphics[width=1\linewidth]{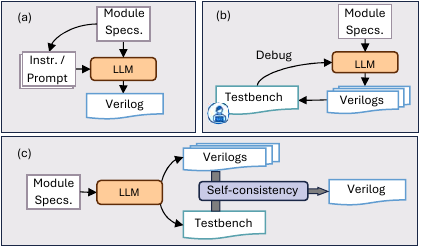}
    \caption{Current training-free approaches: (a) paradigm/prompt engineering, (b) golden-testbench feedback, (c) self-consistency.}
    \label{fig:training-free approaches}
    \vspace{-0.6cm}
\end{figure}

Transformer-based LLMs have revolutionized software code generation, with models like Codex~\cite{chen_evaluating_2021} and AlphaCode~\cite{li_competition-level_2022} achieving impressive results. Inspired by this success, researchers have begun exploring the use of LLMs for Verilog generation. 
However, Verilog generation performance is generally weaker than software code generation, even with fine-tuning on hardware data. This is primarily due to the scarcity of domain-specific data~\cite{pan_survey_2025}. This gap has motivated various training-free strategies to improve LLM-generated Verilog.

training-free approaches aim to improve generation through workflows and post-hoc analysis, as shown in Fig.~\ref{fig:training-free approaches}. Some approaches utilizes prompt engineering to break down generation tasks into pre-defined steps~\cite{sun_paradigm-based_2025}. Another common strategy is testbench-based feedback, where generated Verilog candidates are tested against either golden testbenches or compilation logs. While effective, these methods require human-written testbenches. 

To avoid reliance on golden testbenches, agentic frameworks have been proposed. Among them, VRank introduces a fully automated ranking mechanism. It samples multiple candidate Verilog codes and evaluates them using self-consistency principles. Instead of trusting testbench correctness, VRank observes behavioral agreement across samples and selects the candidate that behaves most consistently across test scenarios.
VRank demonstrates significant improvements in pass@1 accuracy without human intervention. However, it also has key limitations. First, in the pre-ranking stage, it lacks mechanisms to guide the LLM to generate better candidates or verify sample validity before selection. Second, in the post-ranking stage, it relies solely on majority voting and does not explore further enhancement of the selected code. Additionally, it randomly selects from the top-ranked cluster, ignoring potential under-represented edge cases or ambiguous behaviors.

\subsection{Reasoning LLMs and their Potential in Verilog generation}

With the advent of reasoning LLMs such as OpenAI's o1~\cite{openai_openai_2024} and deepseek-r1~\cite{deepseek-ai_deepseek-r1_2025}, test-time scaling has emerged as a promising paradigm. The test-time scaling dynamically increases the reasoning time and reasoning tokens during inference. These reasoning models are trained to have longer steps that include reasoning, just like Chain-of-Thoughts, and are referred to as reasoning LLMs. They have shown great success in math, programming, and logic tasks~\cite{li_system_2025}.

However, reasoning models also suffer from issues as underthinking~\cite{wang_thoughts_2025}, where the model begins to follow a valid reasoning path but prematurely halts or diverges before reaching a solution. This causes a phenomenon where, for the same task, a longer reasoning from the same model is more likely to be incorrect. Laconic decode~\cite{dimakis_laconic_nodate} strategy is proposed to address this problem, prompting LLM for the same math task five times and picking the shortest answer as the final answer.

In Verilog generation, this problem is more complex. Reasoning can lose focus in the process of Verilog generation, leading to poor code quality. We observe that some models can have no in-depth analysis at all, resulting in overly short and wrong reasoning processes. This makes laconic decoding inappropriate with such models in Verilog code generation. 

Given these observations, we enhance Verilog generation by explicitly incorporating reasoning during both candidate generation and refinement. Our proposed framework, VFocus, addresses the limitations of prior agentic methods by deploying reasoning LLMs, ensuring that good samples are obtained by reasoning. After ranking, we further enhance the accuracy by focusing the model on possible inconsistencies it finds in generated codes, and driving it to generate a second round of enhanced code based on its findings.

\begin{figure*}[ht]
\centering
\includegraphics[width=1\linewidth]{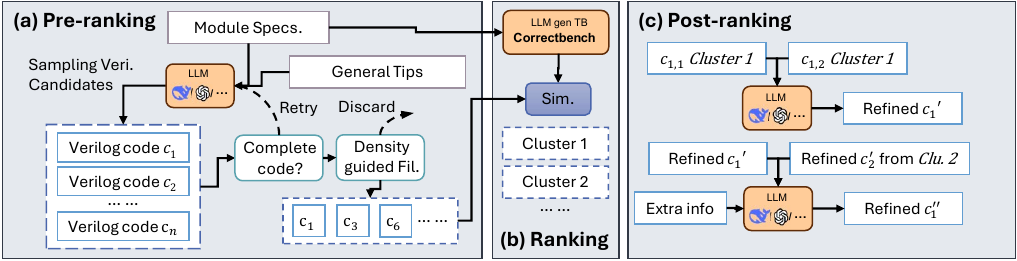} 
\caption{Overall framework of VFocus.} 
\label{fig:overall_framework} 
\vspace{-0.6cm}
\end{figure*}

\section{Proposed Framework}
\label{sec:Proposed Framework}

In VFocus, a reasoning-augmented framework for Verilog code generation is proposed. The input of VFocus is the module specification and the output is the generated Verilog code.
As shown in Fig.~\ref{fig:overall_framework}, VFocus consists of three stages: \textit{pre-ranking sampling and filtering}, \textit{ranking by simulation consistency}, and \textit{post-ranking refinement through reasoning}.
Unlike prior methods that passively rank candidates, VFocus actively sharpens the reasoning focus of LLMs before and after selection, improving both the quality of samples and their correctness.

\subsection{Pre-Ranking Sampling and Filtering: Focused Candidate Preparation}


Given a natural language module specification, we first sample multiple Verilog code candidates using a reasoning LLM. Prompt engineering is applied at this stage. We provide the LLM with general guidelines~\cite{ho_verilogcoder_2024} as well as highlight typical mistakes~\cite{zhao_mage_2024} it tends to make during Verilog generation.

To mitigate syntax errors, we introduce a retry mechanism. If a sampled candidate is syntactically incomplete or invalid, we retry sampling up to a small limit (set to five attempts in this framework), with an increasing delay between retries.

In addition to ensuring syntactic validity, VFocus applies a \textit{Density-guided Filtering} strategy to improve sample quality before ranking. Inspired by observations of reasoning LLM behavior, we measure the token length of each candidate’s reasoning trace and retain only those falling within a ``reasoning sweet spot''—avoiding both too-short (indicative of negligent reasoning) and overly long (suggesting overthinking or deviation) candidates. Specifically, for each sampled candidate $i$, we retain it if its token length $L_i$ satisfies:

\begin{equation}
L_{\text{min}} < L_i < L_{\text{max}}
\end{equation}
where $L_{\text{min}}$ and $L_{\text{max}}$ are empirically determined lower and upper bounds, respectively. Different models may adopt different $(L_{\text{min}}, L_{\text{max}})$ values. We will discuss the choice of these thresholds further in the experiment section.
This filtering ensures that downstream ranking operates on a set of higher-quality, focused samples, ultimately leading to better final candidate selection.

\subsection{Ranking Stage: Simulation-Based Self-Consistency Selection}
For each candidate, we automatically generate a lightweight testbench consisting of multiple test cases to enable behavioral simulation. Each test case corresponds to either an input combination for a combinational circuit or a checkpoint in a sequential circuit. Notably, the testbench does not check outputs explicitly; instead, it prints all relevant outputs for each test case. Testbenches are generated using the CorrectBench~\cite{qiu_correctbench_2024} framework, and further enhanced by reasoning LLMs to ensure that, for sequential circuits, all information related to the current checkpoint is printed out for verification.

After pre-ranking filtering, VFocus simulates each candidate against the generated testbench and collects the resulting output traces. Following the principle of self-consistency~\cite{zhao_codev_2024}, candidates are clustered based on strict behavioral agreement across all test scenarios.

We define the ranking score $R(c)$ of a candidate $c$ as:

\begin{equation} \label{equ:MBR}
R(c) = n - \sum_{c' \in \mathcal{C}} \ell_{\text{strict}}(c, c'), 
\end{equation}

\begin{equation} \label{equ:loss}
\quad \ell_{\text{strict}}(c, c') = \max_{t\in T} \mathbf{1}[c(t)\neq c'(t)]
\end{equation}
where $\mathcal{C}$ denotes the candidate set, $n = |\mathcal{C}|$ is the number of candidates, $T$ represents the set of test cases, and $\mathbf{1}[\cdot]$ is the indicator function.

The reward favors larger clusters, under the assumption that majority consistency correlates with higher correctness. This stage follows the simulation-based ranking mechanism established by VRank~\cite{zhao_vrank_2025}, but benefits from the improved candidate set provided by our pre-ranking filtering.

\subsection{Post-Ranking Refinement: Inconsistency Mining and Reasoning-Enhanced Correction}

While the majority clustering improves robustness, it can still overlook subtle inconsistencies or edge case failures. To address this, VFocus introduces a \textit{post-ranking refinement}. The intuition behind the stage is to repeatedly select two possibly different implementations and try to resolve the inconsistency by reasoning. We employ two strategies in finding inconsistency, namely intra-cluster and inter-cluster inconsistency.

\textbf{Intra-cluster inconsistency}: This step is designed to compensate for the under-representation problem of test cases. Due to our imperfect testbenches that LLM generated, sometimes correct codes and incorrect codes may result in the same cluster. Therefore, in this step, we sample two codes from each top cluster, combined with the module specification, and let LLMs try to find inconsistencies and write a better code for this cluster.

\textbf{Inter-cluster inconsistency}: We analyze simulation outputs from top clusters to identify test scenarios where candidate outputs disagree. These divergences reveal potential areas of behavioral uncertainty. For tasks with a simple description, like waveform description or kmaps, we prompt the LLM with the module specification, the testbench, and the conflicting test input, asking it to reason explicitly about the expected output behavior. For tasks with descriptions of behavior or for tasks whose simulation outputs have too many bits and are non-trivial to reason, we resolve this inconsistency in the same way as intra-cluster inconsistency.

Based on the reasoning outputs, we guide the LLM to generate better code. This step focuses the model’s attention precisely on problematic logic paths, driving targeted correction rather than random resampling.

If no significant inconsistencies are found (an overwhelmingly large cluster consisting of 90\% code candidates), VFocus employs an \textit{early exit} to avoid unnecessary computation. We only keep the intra-cluster resolution and skip the inter-cluster resolution.

Through this reasoning-focused refinement, VFocus enhances Verilog generation quality beyond simple majority voting, effectively combining simulation feedback and structured logical reasoning.

\section{Experimental Setup}
\label{sec:ExperimentalSetup}

\subsubsection{Verilog Generation Benchmark}
We evaluated the performance of our VFocus framework using the VerilogEval-Human benchmark~\cite{liu_verilogeval_2023}, which consists of 156 manually designed Verilog generation tasks. To validate functionality, we simulated the LLM-generated modules and the provided reference testbenches (used only to verify the final selected candidates) using the Icarus Verilog simulator~\cite{williams_steveicarusiverilog_2024}.
                    
\subsubsection{Evaluation Metrics}
Our primary evaluation metric is pass@k, as defined in~\cite{chen_evaluating_2021}, which quantifies the probability that at least one of the top-$k$ generated candidates passes the verification testbench. We compare our method with the random pick baseline and self-consistency method~\cite{zhao_vrank_2025}. Random pick baseline is given by the formulation below:

\begin{equation}
\text{pass@k} := \mathbb{E}_{\text{Problems}} \left[ 1 - \frac{\binom{n-c}{k}}{\binom{n}{k}} \right]
\end{equation}
where \(n\) represents the number of sampled candidates, and \(c\) is the count of correct ones among them. For a fair comparison, we set $n=50$ in all experiments. To reduce variance due to stochastic generation, each experiment was repeated across 5 runs.

To demonstrate the versatility of VFocus, we tested it across a diverse set of language models, covering a range of model sizes, general versus domain-specific pretraining, and both open- and closed-source LLMs.

Our code generation of open-sourced models was carried via online API. Deepseek-r1 provided by the DeepSeek platform, and QwQ-32b on the DeepInfra platform, and o3-mini on the OpenAI API. All code candidates are generated using the model's default recommended temperature setting. Simulations are conducted on a server equipped with dual Xeon Gold 6126 CPUs and 280 GB of RAM.

\section{Experimental Results}
\label{sec:ExperimentalResults}

This section presents an evaluation of the VFocus framework on the VerilogEval-Human benchmark. We assessed its performance across multiple reasoning large language models, including one proprietary model (OpenAI's o3-mini) and two open-source models (Deepseek-R1 and QwQ-32B). The experiments aim to address the following research questions (RQs):
    \textbf{RQ1:} Why is Density-guided Filtering effective?
    \textbf{RQ2:} Does VFocus system further improve Verilog generation quality?
    \textbf{RQ3:} Does VFocus system remains its supremacy in different sample sizes?

\subsection{RQ1: Output pass rate possibility over reasoning length}
\label{sec:rq1_Density}
The first experiment validates the effectiveness of Density-guided Filtering: \textit{correct solutions show similar reasoning pattern, resulting in a similar relative reasoning length} among generated code candidates, while incorrect solutions are more diversely distributed. We analyzed 50 Verilog codes generated by three different models on 156 problems, making it a total of 50*156=7,800 samples. Due to the different complexity of the problems, each graph is normalized to the same 0-1 range for the shortest reasoning to the longest reasoning token numbers in each problem. Note that there are chances that after 5 retries, the code provided is still syntactically incomplete, or the thinking process is missing. Such samples are removed from the graph.

\begin{figure*}[tb]
    \centering
    \begin{subfigure}[b]{0.48\linewidth}%
        \includegraphics[width=\linewidth]{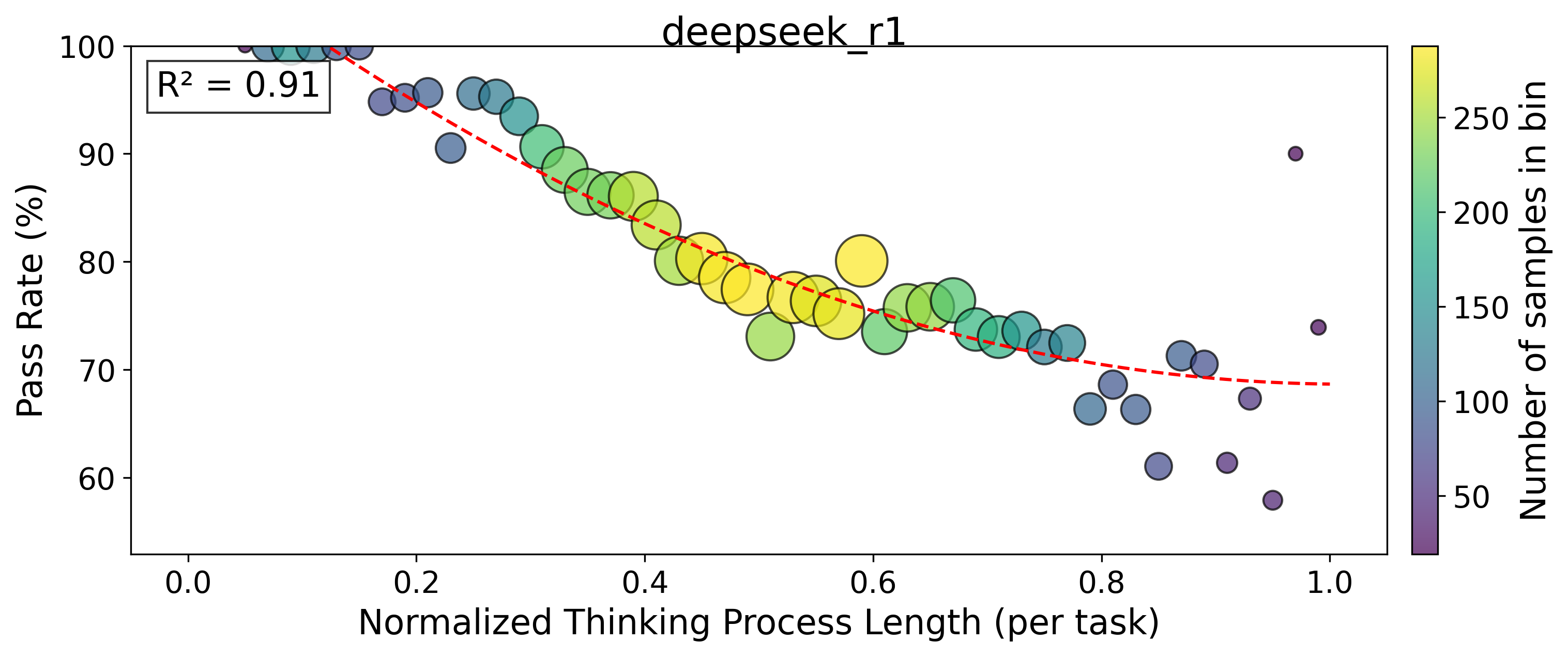}%
        \caption{Deepseek-r1}%
        \label{fig:subfig1}%
    \end{subfigure}%
    \begin{subfigure}[b]{0.48\linewidth}%
        \includegraphics[width=\linewidth]{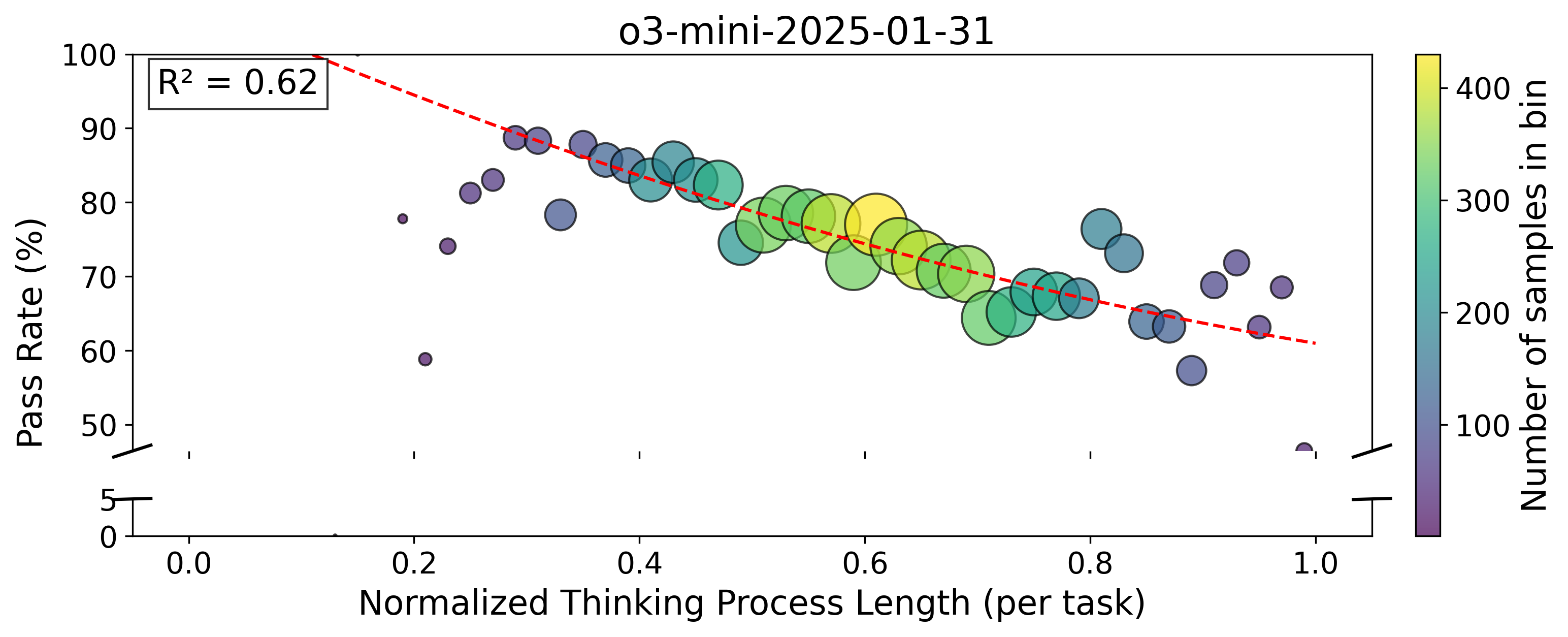}%
        \caption{o3-mini-high}%
        \label{fig:subfig2}%
    \end{subfigure}
    
    \begin{subfigure}[b]{0.48\linewidth}%
        \includegraphics[width=\linewidth]{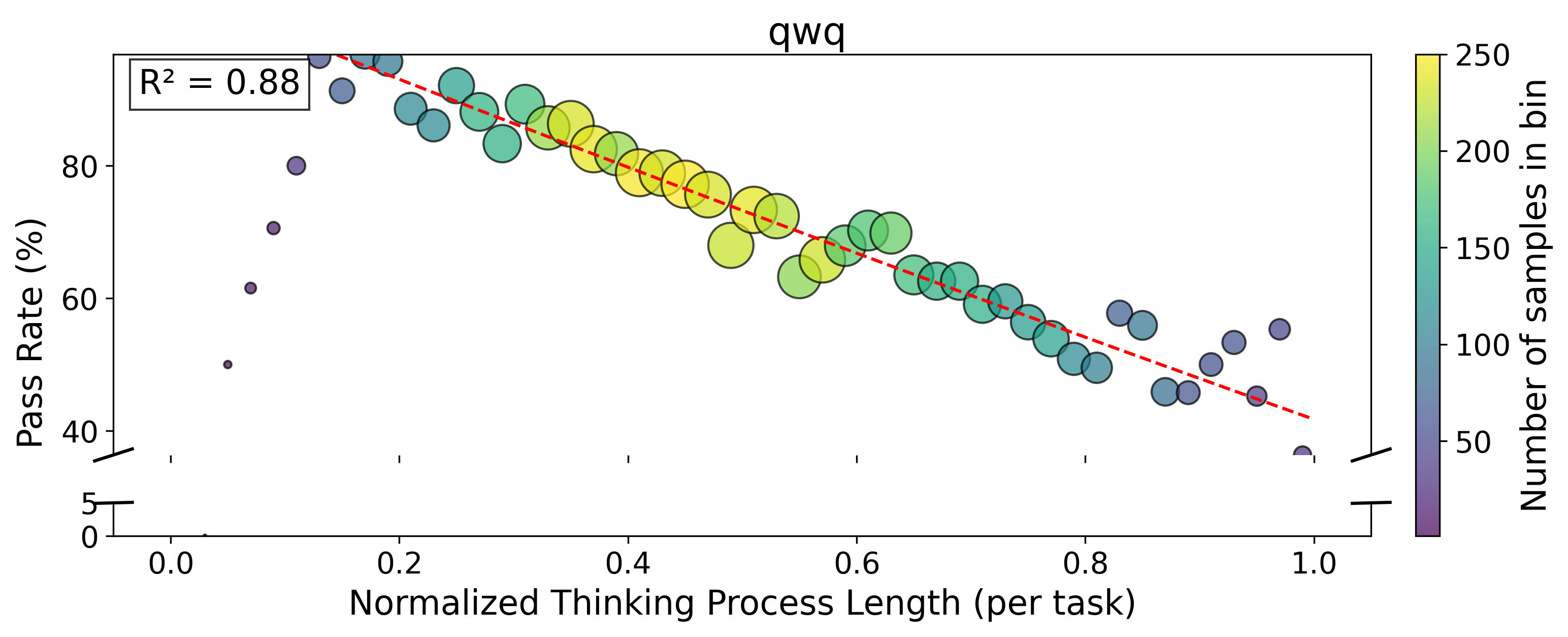}%
        \caption{QwQ-32B}%
        \label{fig:subfig3}%
    \end{subfigure}%
    \begin{subfigure}[b]{0.48\linewidth}%
        \includegraphics[width=\linewidth]{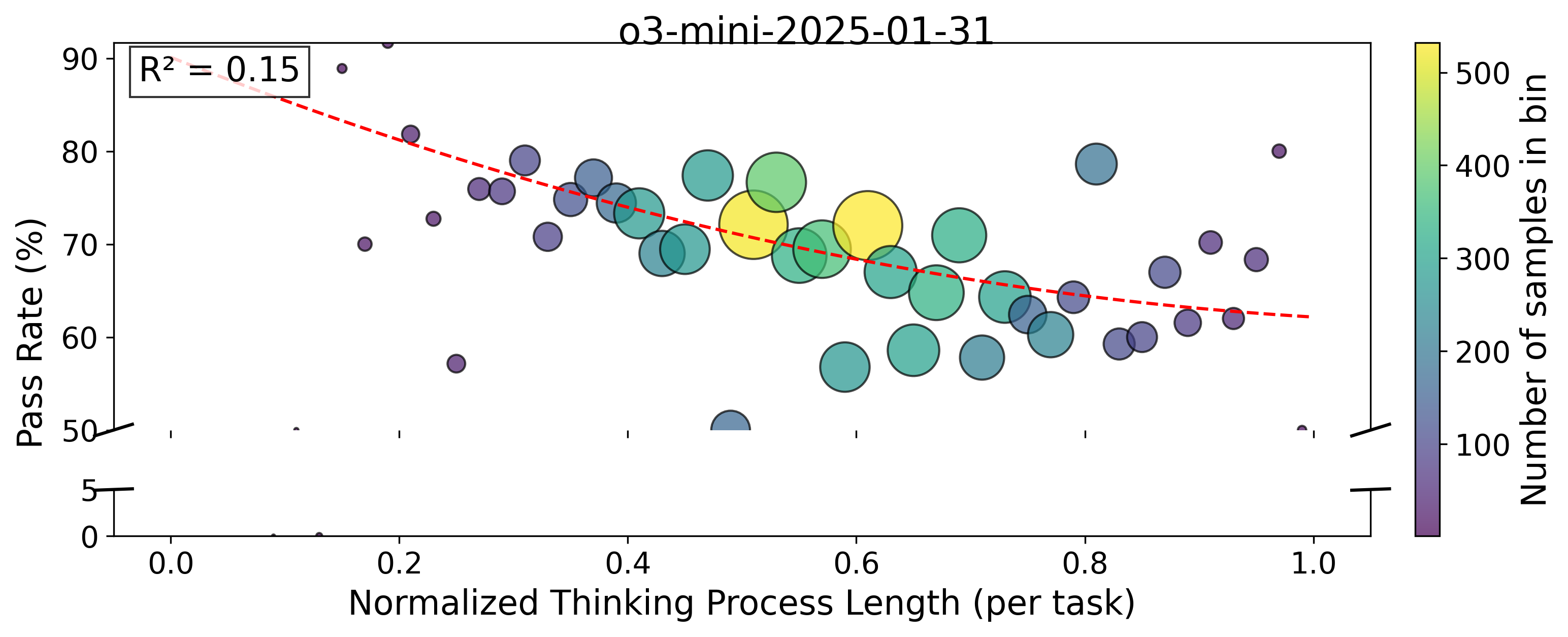}%
        \caption{o3-mini-medium}%
        \label{fig:subfig4}%
    \end{subfigure}%
    \caption{Functional correctness as number of samples increases across different models}%
    \label{fig:sample_length}%
    \vspace{-0.4cm}
\end{figure*}

As shown in Fig. \ref{fig:sample_length} (a), (b) and (c), we draw the pass rate of samples in different models and draw a quadratic trend line to show the approximate trend. The circle in the graph shows how many samples fall into that normalized reasoning length. A clear trend can be seen as the reasoning length goes up, the pass rate is notably decreasing. Another noticeable effect is observed at both qwq and o3-mini-high, that the pass rate also decreases when the reasoning length is too short. This explains the fact why laconic decoding is not effective and stresses the importance of combining the Density-based filtering, syntax completion check, and self-consistency. Based on the observation, we select $L_{\text{max}}$ as the 25\% longest length, and set $L_{\text{min}}$ as the 10\% lowest for both qwq and o3-mini-high, and 0 for deepseek.

An extra observation is that o3-mini-medium in Fig. \ref{fig:sample_length} (d) does not show a similar trend. Unfortunately, due to the lack of detail in its model, and the absence of a reasoning process (OpenAI models do not show any original reasoning text), we can only infer that the token limit they impose after training the model greatly interferes with the model's behavior. Therefore, o3-mini with limited reasoning length may not be suitable for our method of Density-guided Filtering. On the other hand, o3-mini-high, the same model without any inference-time modification, works well with our method.

\begin{table*}[tb]
\centering
\begin{threeparttable}
\caption{Comparison of the proposed framework with direct generation baseline}
\label{table:main_result}
\begin{tabular}{l c |c c c|c|c c}
\toprule[1.2pt]
\multirow{2}{*}{\textbf{Model}} & \multirow{2}{*}{\textbf{Dataset}} & \multicolumn{3}{c|}{\textbf{Baseline}} & \multicolumn{3}{c}{\textbf{Framework Pass@1(Pass@1 increase)} } \\ 
\cmidrule{3-8}
& & \textbf{Pass@1} & \textbf{Pass@2} & \textbf{Pass@3} & \textbf{VRank} & \textbf{Pre+VRank} & \textbf{VFocus} \\ 
\midrule[1.2pt]
Deepseek-R1 & \multirow{3}{*}{Human} & 66.0\% & 70.9\% & 72.9\% & 79.2\% (+13.2\%) & 84.7\%(+18.7\%) & 87.0\%(+21.0\%) \\ 
o3-mini     &                        & 65.3\% & 70.4\% & 72.4\% & 77.4\%(+12.1\%) & 84.2\%(+18.9\%) & 85.6\%(+20.3\%) \\ 
QwQ-32B     &                        & 51.7\% & 58.1\% & 61.1\% & 69.1\%(+17.4\%) & 74.4\%(+22.7\%) & 77.1\%(+25.4\%) \\ 
\midrule[1.2pt]
Deepseek-R1 & \multirow{3}{*}{CMB(81)} & 83.1\% & 87.7\% & 89.4\% & 94.9\%(+11.8\%) & 94.8\%(+11.7\%) & 95.0\%(+11.9\%) \\ 
o3-mini     &                          & 78.6\% & 83.7\% & 85.3\% & 88.9\%(+10.3\%) & 94.1\%(+15.5\%) & 94.3\%(+15.7\%) \\ 
QwQ-32B     &                          & 70.5\% & 77.3\% & 80.0\% & 88.2\%(+17.7\%) & 93.6\%(+23.1\%) & 93.3\%(+22.8\%) \\ 
\midrule[1.2pt]
Deepseek-R1 & \multirow{3}{*}{SEQ(75)} & 47.6\% & 52.3\% & 55.2\% & 62.2\%(+14.6\%) & 72.4\%(+24.8\%) & 78.5\%(+30.9\%) \\ 
o3-mini     &                          & 50.9\% & 56.0\% & 58.4\% & 63.8\%(+12.9\%) & 73.4\%(+22.5\%) & 76.2\%(+25.3\%) \\ 
QwQ-32B     &                          & 31.5\% & 37.4\% & 40.6\% & 48.6\%(+17.1\%) & 53.7\%(+22.2\%) & 59.6\%(+28.1\%) \\ 
\bottomrule[1.2pt]
\end{tabular}

\end{threeparttable}
\vspace{-0.5cm}
\end{table*}

\subsection{RQ2: Accuracy Improvement Across Models}
\label{sec:rq2_accuracy_improvement}

The performance of VFocus on different LLMs is presented in Table \ref{table:main_result}. The reported baseline is to use the original prompt of VerilogEval and random sampling, which represents the original default method of generating Verilog from a specification. We report the pass@1 scores for both the baseline, self-consistency method (VRank), the pass@1 score after pre-ranking improvement, and VFocus.

Our experimental results demonstrate that VFocus successfully further improves Verilog correctness on reasoning models, consistently improves accuracy across all tested reasoning models, including both closed-source and open-source models. The first observation we have is the reasoning LLMs' supremacy over non-reasoning LLMs. Compared to the state-of-the-art high of baseline (random pick) achieved by GPT-4o (57.4\% pass@1 reported by~\cite{zhao_vrank_2025}), the baseline achieved by all three models tested surpasses the GPT-4o by a large margin.

While self-consistency methods remain effective on reasoning LLMs, our framework further enhance the performance of pass@1 considerably in all three models, achieving an accuracy as high as 84.7\% with pre-ranking refinement on deepseek-R1. We also evaluate how post-ranking refinement further enhances the pass@1 accuracy of our framework. We applied both intra-cluster and inter-cluster reasoning to all three models, with the early-exit strategy applied. All three models show steady improvement on Pass@1 accuracy, as shown in the Table. ~\ref{table:main_result}.

Notably, our VFocus framework shows more improvement on sequential circuits, demonstrating +30.9\% improvement compared to the baseline of Deepseek-R1, and a +16.3\% improvement over the self-consistency method of VRank. In contrast, the improvements on combinational circuits are more modest, primarily due to their already high baseline accuracy.

\subsection{RQ3: Impact of Sample Size on Performance}
\label{sec:rq3_sample_size}

\begin{figure*}[tb]
    \centering
    \includegraphics[width=1\linewidth]{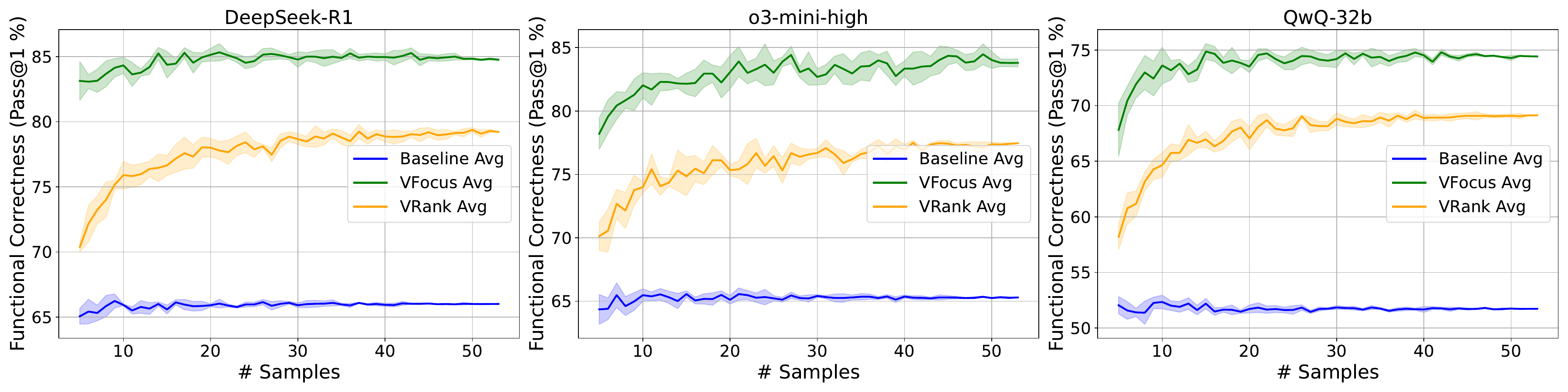}
    \caption{Functional correctness increase as \# Samples increase}
    \label{fig:sample_num}
    \vspace{-0.6cm}
\end{figure*}
We assessed the Pass@1 accuracy of VFocus across various LLMs by varying the number of generated code candidates from 5 to 50, enabling a comparative analysis against both the baseline and VRank. Each experiment was repeated 10 times, and we report the average accuracy (as lines) along with the standard deviation (as shaded regions) in Fig. \ref{fig:sample_num}. 
As shown in the graph, VFocus steadily outperforms both VRank and baseline. Due to the high repetitive cost of post-ranking refinement, the graph does not include the performance of the post-ranking stage but only reports the ranking result. As shown in the graph, the pre-ranking strategy is highly effective, making our framework steadily outperform the baseline and self-consistency method by a large margin. The margin is even larger in a smaller sample size. This is because the self-consistency method requires high-quality samples, and in a smaller sample size, if we have many invalid samples or low-quality samples, this will greatly degrade the performance.

\section{Conclusion}
\label{sec:conclusion}
In this paper, we introduced VFocus, a framework that enhances Verilog generation by sharpening the focus of reasoning LLMs on critical decision points throughout the code generation process. The pre-ranking stage employs a retry mechanism and Density-guided Filtering to retain candidates within a reasoning "sweet spot", balancing depth and brevity to maximize functional correctness. These strategies ensure that candidates are neither under-reasoned nor over-complicated. The post-ranking refinement stage directs the model’s attention to inconsistencies between the top candidates, enabling targeted debugging through scenario-specific reasoning prompts. Experiments on VerilogEval-Human demonstrate that VFocus achieves significant pass@1 improvements over baseline methods across diverse reasoning LLMs, including both closed-source and open-source models. Notably, VFocus operates autonomously without relying on human-written testbenches, making it scalable for real-world hardware design tasks. Our work highlights the importance of aligning LLM reasoning focus with code generation subtleties in hardware languages, paving the way for more reliable AI-driven design automation tools.

\section*{Acknowledgment}
This work is supported by the Deutsche Forschungsgemeinschaft
(DFG, German Research Foundation) -Project-ID 504518248.



\printbibliography

\end{document}